\documentclass[aps,preprint,a4paper,showpacs,showkeys,superscriptaddress]{revtex4-1}
\usepackage{latexsym}
\usepackage{amsmath,amssymb}
\usepackage{graphicx}
\usepackage{subfigure}
\usepackage{xcolor}
\usepackage{cancel}

\usepackage{hyperref}     % color hypertext for pdflatex
\hypersetup{colorlinks,%
  citecolor=blue,%
  linkcolor=cyan,%
  pdftex}

\begin{document}

%\preprint{arXiv:yymm.nnnn [gr-qc]}

\title{Effective Tolman temperature induced by trace anomaly}

\author{Myungseok Eune}%
\email[]{eunems@smu.ac.kr}%
\affiliation{Department of Civil Engineering, Sangmyung
  University, Cheonan, 31066, Republic of Korea}%

\author{Yongwan Gim}%
\email[]{yongwan89@sogang.ac.kr}%
\affiliation{Department of Physics, Sogang University, Seoul, 04107,
  Republic of Korea}%
\affiliation{Research Institute for Basic Science, Sogang University,
  Seoul, 04107, Republic of Korea} %

\author{Wontae Kim}%
\email[]{wtkim@sogang.ac.kr}%
\affiliation{Department of Physics, Sogang University, Seoul, 04107,
  Republic of Korea}%

\date{\today}

\begin{abstract}
  Despite the finiteness of stress tensor for a scalar field on the four-dimensional
  Schwarzschild black hole in the Israel-Hartle-Hawking vacuum,
  the Tolman temperature in thermal equilibrium is certainly
  divergent on the horizon due to the infinite blueshift of the Hawking temperature.
  The origin of this conflict is due to the fact
  that the conventional Tolman temperature was based on the assumption of a traceless stress tensor, 
which is, however, incompatible with the presence of the trace anomaly responsible
  for the Hawking radiation.
  Here, we present an effective Tolman temperature which is compatible with the presence of the trace anomaly by
  using the modified Stefan-Boltzmann law.
  Eventually, the effective Tolman temperature turns out to be finite everywhere outside the horizon,
  and so there does not appear
  infinite blueshift of the Hawking temperature at the event horizon any more. In particular, it is
 vanishing on the horizon, so that the equivalence principle is exactly recovered at the horizon.
  \end{abstract}

% \pacs{04.70.Dy, 04.62.+v, 04.60.Kz }


\maketitle
%%%%%%%%%%%%%%
%% RevTeX Style End %%
%%%%%%%%%%%%%%

%\newcommand{\lp}{\ell_P}

\section{Introduction}
\label{sec:intro}
%%%%%%%%%%%%something introduction%%%%%%%%%%%%%%%%%%%%%%%%%%%%%%%%%%%%%%%%%%

 A quantum black hole~\cite{Hawking:1974sw, Hawking:1976ra} in the Israel-Hartle-Hawking vacuum
 \cite{Hartle:1976tp,Israel:1976ur} could be characterized by
 the Hawking temperature $T_{{\rm H}}$ which is given by
 the surface gravity. The local temperature in a proper frame as the Tolman temperature
 can be defined  in the form of the blue-shifted Hawking temperature as
 \cite{Tolman:1930zza, Tolman:1930ona}
 \begin{equation}\label{localtem}
 T_{{\rm loc}} = \frac{T_{\rm H}}{\sqrt{-g_{00}(r)}},
\end{equation}
which is infinite at the horizon due to the infinite blueshift of the Hawking temperature,
though it reduces to the Hawking temperature at infinity.

On the other hand, the renormalized stress tensor
for a conformal scalar field could be finite
 on the background of the
 Schwarzschild black hole \cite{Page:1982fm}.
 At infinity, the proper energy density $\rho$
 is positive finite, which is consistent with
the Stefan-Boltzmann law as $\rho =\sigma T^4_{\rm H}$, where
$\sigma =\pi^2/30$.
If one considered a motion of an inertial
 observer \cite{Page:1982fm, Ford:1993bw, Ford:2009vz,Freivogel:2014dca}, the negative proper energy density could be found
 near the horizon in various vacua and
its role was also discussed in
 connection with the information loss paradox \cite{Freivogel:2014dca}.
However, it might be interesting
to note that the local temperature \eqref{localtem} is infinite
at the horizon, although the proper energy density
at the horizon $r_{\rm H}$ is negative finite
as $\rho(r_{\rm H}) =-12 \sigma T^4_{\rm H}$ as seen in Ref. \cite{Page:1982fm}.

 Now, it appears to be puzzling in that
 the Tolman temperature at the horizon is positively divergent despite
 the negative finite energy density there.
  More worse, the energy density happens to vanish at a certain point outside the
 horizon \cite{Page:1982fm}, but the local temperature \eqref{localtem} is positive finite at that point.
 In these regards, the Tolman temperature runs contrary to
 the finite renormalized stress tensor, which certainly requires that
 the Stefan-Boltzmann law to relate the stress tensor to the proper temperature
 should be appropriately modified in such a way that they are compatible each other.

 To resolve the above conflict between the finiteness of the renormalized stress tensor
 and the divergent behavior of the proper temperature,
 it is worth noting that the usual Tolman temperature rests upon the traceless stress tensor;
 however, the trace of the renormalized stress tensor is actually not traceless because of
 the trace anomaly.
  So we should find a modified Stefan-Boltzmann law
  in order to get the proper temperature commensurate with
  the finite renormalized stress tensor.
 In fact, this was successfully realized in the two-dimensional case
 where the stress tensor was perfect fluid~\cite{Gim:2015era}.
 In this work, we would like to extend the above issue to the case of
 the four-dimensional more realistic Schwarzschild black hole, where
 the renormalized stress tensor is no more isotropic.

Using the exact thermal stress
tensor calculated in Ref.~\cite{Page:1982fm},
we solve the covariant conservation law and
the equation for the trace anomaly, and then obtain the proper quantities such as the proper energy
density and pressures written explicitly in terms of the trace anomaly in Sec.~\ref{sec:4D}.
In Sec.~\ref{sec:4}, we derive the effective Tolman temperature from the modified Stefan-Boltzmann law
based on thermodynamic analysis.
It shows that the effective Tolman temperature exactly reproduces the
Hawking temperature at infinity, but it has a maximum at a finite
distance outside the horizon and eventually it is vanishing rather than divergent on the horizon.
Finally, conclusion and discussion are given in Sec.~\ref{sec:Dis}.

\section{Proper quantities in terms of trace anomaly}
\label{sec:4D}
% \subsection{Derivation of generalized temperature}
% \label{subsec:4DTolman}

We start with a four-dimensional
Schwarzschild black hole governed by the static line element as
\begin{equation}
  \label{4Dmetric}
  ds^2 =- f(r)dt^2 + \frac{1}{f(r)} dr^2  + r^2 (d \theta ^2 +\sin^2 \theta d\phi^2),
\end{equation}
where the metric function is $f(r)=1-2GM/r$.
The renormalized stress tensor for a conformal scalar field on
the Schwarzschild black hole was obtained in the Israel-Hartle-Hawking vacuum  \cite{Hartle:1976tp,Israel:1976ur} by using the
Gaussian approximation as~\cite{Page:1982fm}
\begin{align}
  T^\mu_\nu &= \frac{\pi^2}{90} \left(\frac{1}{8\pi M}\right)^4 \left[
              \frac{1-(4-\frac{6M}{r})^2(\frac{2M}{r})^6}{(1-\frac{2M}{r})^2}
              (\delta^\mu_\nu-4\delta^\mu_0\delta^0_\nu) + 24
              \left(\frac{2M}{r}\right)^6 (3\delta^\mu_0\delta^0_\nu +
              \delta^\mu_1\delta^1_\nu) \right], \label{eq:T:ij:Page}
\end{align}
where it is finite everywhere.

On general grounds, the trace anomaly
can be written in the form of curvature invariants as
\begin{equation}
T^\mu_\mu = \alpha \left( \mathcal{F}+ \frac{2}{3} \Box R \right) + \beta \mathcal{G},
\end{equation}
where $\mathcal{F} = R^{\mu\nu\rho\sigma}R_{\mu\nu\rho\sigma} - 2 R^{\mu\nu}
R_{\mu\nu} + R^2/3$ and $ \mathcal{G} = R^{\mu\nu\rho\sigma}R_{\mu\nu\rho\sigma} - 4
R^{\mu\nu} R_{\mu\nu} + R^2$~\cite{Deser:1976yx,
  Duff:1977ay, Birrell:1982ix, Deser:1993yx,Duff:1993wm}.
There have been
a lot of applications of trace anomalies to Hawking radiation and
black hole thermodynamics in wide variety of cases of interest \cite{Elizalde:1992mt, Elizalde:1992zm, Nojiri:1998ue, Nojiri:1998ph, Nojiri:2000ja,
  Elizalde:1999dw, Burinskii:2001bq, Cai:2009ua, Kawai:2014afa, Vieira:2015oka,
  Kawai:2015uya, Kawai:2017txu}. The coefficients $\alpha$ and $\beta$ are related
to the number of conformal fields such as real scalar fields
$N_{\rm S}$, Dirac (fermion) fields $N_{\rm F}$, and vector fields
$N_{\rm V}$, such that they are fixed as
$\alpha = (120 (4\pi)^2)^{-1} (N_{\rm S} + 6 N_{\rm F} + 12 N_{\rm
  V})$ and $\beta = - (360 (4\pi)^2)^{-1} (N_{\rm S} + 11 N_{\rm F} + 62 N_{\rm
  V})$. For the Ricci flat spacetime with a single conformal scalar field, the trace
anomaly reduces to
\begin{align}
  T^\mu_\mu = \frac{1}{2880\pi^2}R^{\mu\nu\rho\sigma}
  R_{\mu\nu\rho\sigma}= \frac{M^2}{60\pi^2 r^6}, \label{eq:4D:Sch:anomaly}
\end{align}
and then the trace for the stress tensor~\eqref{eq:T:ij:Page} is exactly
in accord with the conformal
anomaly~\eqref{eq:4D:Sch:anomaly}.

In contrast to the two dimensional case~\cite{Gim:2015era},
the stress tensor appears anisotropic in the spherically symmetric black
hole in four dimensions, and so the form of the stress
tensor~\eqref{eq:T:ij:Page} should be generically written as~\cite{Hawking:1973,
  Lobo:2007zb}
\begin{align}
  T^{\mu\nu} = (\rho+p_t)u^\mu u^\nu +p_t g^{\mu\nu}+(p_r-p_t)
  n_{(r)}^\mu n_{(r)}^\nu.   \label{eq:EM:anistropic}
\end{align}
The proper velocity $u^\mu$ is a timelike unit vector satisfying
$u^\mu u_\mu = -1$, $n^\mu_{(r)}$ is the unit spacelike vector in
the radial direction, and $n_{(\theta)}^\mu$ and $n_{(\phi)}^\mu$
are the unit normal vectors which are orthogonal to $n^\mu_{(r)}$
satisfying $g_{\mu \nu} n_{(i)}^\mu n_{(j)}^\mu = \delta_{ij}$ and
$n_{(i)}^\mu u_\mu=0$ where $i,j = r,\theta, \phi$.  Thus the
spacelike unit normal vectors are determined as
\begin{align}
  n^\mu_{(r)} &=\left(0, \sqrt{f(r)}, 0, 0
                \right), \quad n^\mu_{(\theta)}=\left(0, 0, \frac{1}{r}, 0 \right),
                \quad n^\mu_{(\phi)}=\left( 0, 0, 0,
                \frac{1}{r \sin\theta} \right), \label{n}
\end{align}
with the proper velocity
\begin{align}
  u^\mu = \left(\frac{1}{ \sqrt{f(r)}}, 0, 0, 0\right)   \label{4Dvelocity}
\end{align}
for the frame dropped from rest.  Then, from
Eqs.~\eqref{eq:T:ij:Page}, \eqref{eq:EM:anistropic}, \eqref{n},
and~\eqref{4Dvelocity}, the proper energy density and pressures can be
explicitly calculated by using the following relations,
\begin{align}
  \rho = T_{\mu\nu} u^\mu u^\nu,~p_r = T_{\mu \nu} n_{(r)}^\mu
  n_{(r)}^\nu,~ p_t = T_{\mu \nu} n_{(\theta)}^\mu n_{(\theta)}^\nu =
  T_{\mu \nu} n_{(\phi)}^\mu n_{(\phi)}^\nu, \label{proper}
\end{align}
where the proper flux along $x^i$-direction can also be obtained by using
the relation $\mathcal{F}_i=-T_{\mu\nu}u^\mu n_{(i)}^\nu$ but it
trivially vanishes in thermal equilibrium~\cite{Hartle:1976tp, Israel:1976ur}.

Note that the energy density and pressures are not independent as seen from the
trace relation,
\begin{align}
  T^\mu_\mu = -\rho + p_r + 2 p_t.   \label{4Dtrace}
\end{align}
From Eqs.~\eqref{eq:T:ij:Page}, \eqref{eq:4D:Sch:anomaly},
and~\eqref{proper}, we find an additional relation
\begin{align}
  p_r-p_t = \frac{1}{4} T^\mu_\mu , \label{eq:p.r:p.t}
\end{align}
which characterizes the anisotropy
between the tangential pressure and radial pressure.

Let us now express the proper energy density and pressures formally in
terms of the trace anomaly for our purpose.  From
Eq.~\eqref{eq:EM:anistropic}, the covariant
conservation law for the energy-momentum tensor is rewritten as
\begin{align}
  \partial_r p_r + \frac{2}{r} (p_r-p_t) + \frac{1}{2f} \partial_r f
  (p_r+\rho)  =0.   \label{4Dsource}
\end{align}
Plugging Eqs.~\eqref{4Dtrace} and~\eqref{eq:p.r:p.t}
 into Eq.~\eqref{4Dsource}, one can obtain the simplified form of
\begin{align}
  \partial_r  p_r + \frac{\partial_r f}{2f}p_r = - \left(\frac{1}{2r}
    + \frac{3 \partial_r f}{4f}\right)T^\mu_\mu,   \label{4DDfp}
\end{align}
which can be solved as
\begin{align}
  p_r = \frac{1}{f^2} \left(C_0 + \int^r \frac{f}{4r}(-2f
  +3r \partial_r f)  T^\mu_\mu  dr \right),   \label{4DimprovedPr}
\end{align}
where $C_0$ is an integration constant.
Additionally, from Eqs.~\eqref{4Dtrace} and \eqref{eq:p.r:p.t}, the
tangential pressure and energy density can also be obtained as
\begin{align}
  p_t &= \frac{1}{f^2}\left(C_0-\frac{f^2}{4}T^\mu_\mu+\int^r
        \frac{f}{4r}(-2f +3r \partial_r f)  T^\mu_\mu  dr
        \right),\label{4DimprovedPt} \\
  \rho &= \frac{3}{f^2}\left(C_0-\frac{f^2}{2}T^\mu_\mu+\int^r
         \frac{f}{4r}(-2f +3r \partial_r f)  T^\mu_\mu  dr \right). \label{4Dimprovedrho}
\end{align}
The above proper quantities defined in freely falling frames
were related to the trace anomaly conveniently, which will be used in the next section.
 \section{Effective Tolman temperature}
\label{sec:4}

In this section, we derive the proper temperature for the background
of the four-dimensional Schwarzschild black hole based on
the modified Stefan-Boltzmann law.
First of all, we note that the volume of the system in the radial proper frame can
be changed only along the radial direction on the spherically
symmetric black hole,
and thus obtain the thermodynamic first law written
as
\begin{align}
  dU = T dS - p_r dV  \label{eq:1st.law}
\end{align}
without recourse to the tangential work. From Eq.~\eqref{eq:1st.law}, one
can immediately get
\begin{align}
  \left(\frac{\partial U}{\partial V} \right)_T = T
  \left(\frac{\partial S}{\partial V} \right)_T -
  p_r,   \label{eq:1st.law:another}
\end{align}
and then from the Maxwell relations such as
$(\partial S/\partial V)_T= (\partial p_r /\partial T)_V$,
we obtain
\begin{align}
  \rho =  T \left(\frac{\partial p_r}{\partial T} \right)_V -
  p_r.   \label{eq:eom:therm.}
\end{align}
Using the fact that the trace anomaly is independent of temperature as
$\partial_{T} T^\mu_\mu =0$~\cite{BoschiFilho:1991xz},
from Eqs.~\eqref{4Dtrace} and \eqref{eq:p.r:p.t}, we also obtain
\begin{align}
  \left(\frac{\partial \rho}{\partial T}\right)_V
  =\left(\frac{\partial p_r}{\partial
      T}\right)_V+2\left(\frac{\partial p_t}{\partial
      T}\right)_V, \label{partial p1}
\end{align}
and
\begin{equation}
  \left(\frac{\partial p_r}{\partial T}\right)_V=\left(\frac{\partial
      p_t}{\partial T}\right)_V. \label{partial p2}
\end{equation}
Plugging Eqs.~\eqref{partial p1} and \eqref{partial p2} into
Eq.~\eqref{eq:eom:therm.}, we get
\begin{equation}
 \label{eq:4D:p.r:eq}
 T \left(\frac{\partial \rho}{\partial T}\right)_V-4\rho = \frac{3}{2}T^\mu_\mu,
\end{equation}
which is solved as
\begin{align}
  \rho = 3\gamma T^4 - \frac{3}{8}T^\mu_\mu.\label{4Dr}
\end{align}
From Eqs.~\eqref{4Dtrace} and~\eqref{eq:p.r:p.t}, the radial and
tangential pressure are also derived as
\begin{align}
  p_r &= \gamma T^4 + \frac{3}{8}T^\mu_\mu, \label{4Dpr} \\
  p_t &= \gamma T^4 + \frac{1}{8}T^\mu_\mu, \label{4Dpt}
\end{align}
respectively.  The integration constant $\gamma$ is related to
the Stefan-Boltzmann constant $\sigma$ as $\gamma = \sigma/3=\pi^2/90$~ for a
conformal scalar field~\cite{Christensen:1977jc}.
For the traceless case,
the modified Stefan-Boltzmann law \eqref{4Dr}
simply reduces to the usual one.  The proper energy density in Eq.~\eqref{4Dr} is not
necessarily positive definite thanks to the trace anomaly, so that
the negative energy states are naturally permitted in this extended setting.

From Eqs.~\eqref{4Dr}, \eqref{4Dpr}, and \eqref{4Dpt}, the proper
temperature is obtained as
\begin{align}
  T = \left[\frac{1}{\gamma} \left(p_r-\frac{3}{8} T^\mu_\mu \right)
  \right]^{1/4}  = \left[\frac{1}{\gamma} \left(p_t-\frac{1}{8}
  T^\mu_\mu \right) \right]^{1/4}  = \left[\frac{1}{3\gamma}
  \left(\rho+\frac{3}{8} T^\mu_\mu \right) \right]^{1/4},
\end{align}
and it can be compactly written in terms of the trace anomaly as
\begin{equation}\label{eq:4D:T}
  T =\frac{1}{\gamma^{1/4} \sqrt{f}} \left( C_0 - \frac{3}{8} f^2
    T^\mu_\mu+\int^r \frac{f}{4r}(-2f +3r \partial_r f)  T^\mu_\mu  dr
  \right)^{1/4},
\end{equation}
where we used Eqs.~\eqref{4DimprovedPr}, \eqref{4DimprovedPt}, and
\eqref{4Dimprovedrho}.  In the absence of the trace anomaly, the
proper temperature~\eqref{eq:4D:T} reduces to the usual Tolman
temperature~\cite{Tolman:1930zza, Tolman:1930ona}.
Requiring that the proper temperature~\eqref{eq:4D:T} be coincident with
the Hawking temperature $T_{\rm H}$ at infinity, we can fix the constant
as $C_0=\gamma^{1/4} T_{\rm H}$.

%%%%%%%%%%%%%%%%%%%%%%%%%%%%%
% Fig: %%
%%%%%%%%%%%%%%%%%%%%%%%%%%%%%
\begin{figure}[hpt]
  \begin{center}
  \includegraphics[width=0.7\linewidth]{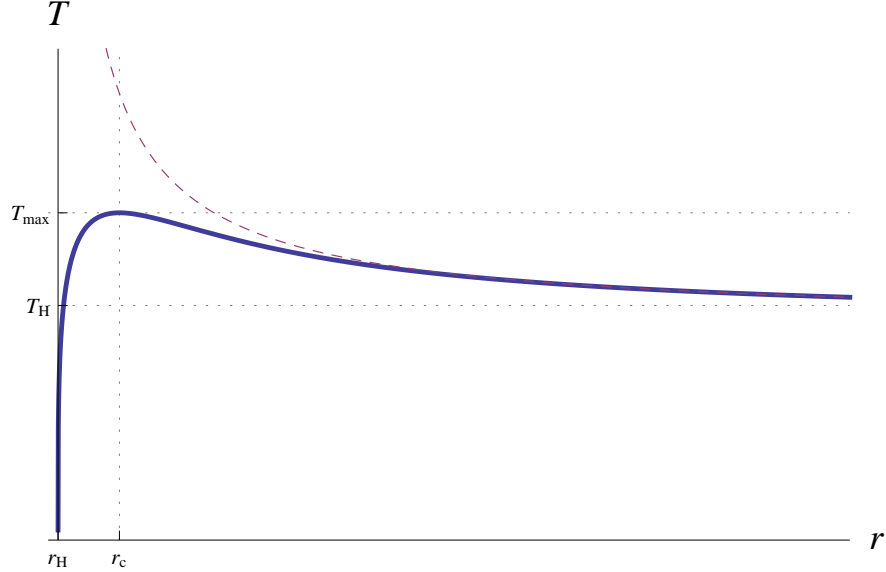}
  \end{center}
  \caption{The dashed curve shows the behavior of the usual Tolman temperature
    of being divergent on the horizon. The solid curve is
    the effective Tolman temperature, which is
    finite everywhere. In particular, it vanishes at the horizon and has
    a maximum $T_{\rm max} \sim 1.51 T_{\rm H}$ at
    $r_c \sim 1.31 r_H$. All the curves approach the Hawking
    temperature at infinity, whereas they are very different from each
    other near the horizon where quantum effects are significant.}
  \label{fig:Tvsr}
\end{figure}
%%%%%%%%%%%%%%%%%%%%%%%%%%%%%

Finally, plugging the trace
anomaly~\eqref{eq:4D:Sch:anomaly} into Eq.~\eqref{eq:4D:T}, we obtain
\begin{align}
  T=\frac{1}{8\pi M \sqrt{f(r)}} \left[ 1 -
  28 \left(\frac{2M}{r}\right)^6 + 48 \left(\frac{2M}{r}\right)^7 - 21
  \left(\frac{2M}{r} \right)^8 \right]^{1/4},
\end{align}
which can be neatly factorized as
\begin{align}
  T = \frac{1}{8\pi M \sqrt{f(r)}}
  &\left[\left(1 - \frac{2M}{r}\right)^2 \left(1 +
    2\left(\frac{2M}{r}\right) + 3\left(\frac{2M}{r}\right)^2
    \right. \right. \notag \\
  & \left. \left. + 4 \left(\frac{2M}{r}\right)^3 +
    5\left( \frac{2M}{r}\right)^4 + 6\left( \frac{2M}{r}\right)^5 -
    21\left( \frac{2M}{r}\right)^6 \right) \right]^{1/4}.
\end{align}
It seems to be interesting to note
that the blueshift factor in the denominator related to the origin of the divergence at the horizon
can be canceled out, so that the
effective Tolman temperature is written as
\begin{align}
  T = \frac{1}{8\pi M}\left[\left(1 - \frac{2M}{r} \right) \sum_{n=1}^6
  \frac{n(n+1)}{2} \left(\frac{2M}{r}\right)^{n-1} \right]^{1/4}. \label{4DnewT}
\end{align}
Thus the redshift factor responsible for the
infinite blueshift of the Hawking temperature on the horizon does not
appear any more in the effective Tolman temperature.
As seen from Fig.~\ref{fig:Tvsr}, the behavior of the temperature~\eqref{4DnewT}
shows that it is finite everywhere and approaches the Hawking
temperature at infinity.  In particular, it is vanishing on the
horizon, so that the freely falling observer from rest does not see any excited particles.
On the contrary to the naively expected divergence from the
usual Tolman temperature at the horizon, the high frequency quanta
could not be found on the horizon, which would be compatible with the result that
the equivalence principle could be recovered at the
horizon~\cite{Singleton:2011vh}.

The divergent
dashed curve near the horizon in Fig.~\ref{fig:Tvsr} could be made finite
by taking into account the quantum effect via the trace anomaly, which is reminiscent of the
vanishing Hawking temperature in the noncommutative Schwarzschild black
hole based on the different assumptions of quantization rules~\cite{Nicolini:2008aj}.
And the proper temperature based on the effective temperature method
is also compatible with the present result in the sense that
the proper temperature vanishes at the horizon \cite{Barbado:2016nfy}.

\section{Conclusion and Discussion}
\label{sec:Dis}
It has been widely believed that the Tolman temperature
is divergent at the horizon due to the infinite blueshift of the Hawking radiation.
However, the usual Stefan-Boltzmann law assuming the traceless stress tensor should be
consistently modified in order to discuss the case where the stress
tensor is no longer traceless in the process of the Hawking radiation.
From the modified Stefan-Boltzmann
law, we obtained the effective Tolman temperature without the redshift factor
related to the origin of the divergence at the horizon, so that it is
finite everywhere outside the black hole horizon.

The intriguing behavior of the effective Tolman temperature on the horizon may be understood
by the Unruh effect~\cite{Unruh:1976db}. The static
metric~\eqref{4Dmetric} near the horizon can be written by the Rindler
metric for a large black hole whose curvature scale is negligible.
The Unruh temperature is
divergent due to the infinite acceleration of the frame where the
fixed detector is very close to the horizon.
So the Unruh temperature is equivalent to the locally
fiducial temperature for the Schwarzschild black hole~\cite{Singleton:2011vh}.
Conversely speaking, based on the equivalence principle,
the Unruh temperature measured by the geodesic detector should vanish on the
horizon since the proper acceleration of the geodesic detector vanishes.
In this regard, it appears natural to
conclude that the freely falling observer from rest does not see any excited
particles on the horizon in thermal equilibrium and thus the effective Tolman
temperature vanishes at the horizon.

 On the other hand, AMPS argument is that the firewall  
 on the horizon should be defined in an
 evaporating black hole rather than the black hole in thermal equilibrium \cite{Almheiri:2012rt}.
 The firewall is certainly characterized by the divergent proper temperature
 in that the average frequency $\omega$ of an excited particle with a
 thermal bath can be identified with the proper
 temperature as $ \omega \sim T$.
 Using the advantage of the effective Tolman temperature, we 
 find the reason why the firewall
 could not exist in thermal equilibrium, that is, 
 from the fact that the redshift factor responsible for the divergence at the horizon
 could be canceled out.

\acknowledgments

We would like to thank Jeong-Hyuck Park and Edwin
J. Son for exciting discussions. W. Kim was supported by
the National Research Foundation of Korea(NRF) grant funded by the
Korea government(MSIP) (2017R1A2B2006159).

%%%%%%%%%%%%%%%%%%%%%%%%%%%%%%%%%%%%%%%%%%%%%%%%
%%%%%%%%%%%%%%%             References         %%%%%%%%%%%%%%%%

%\bibliographystyle{mybib}
%\bibliographystyle{apsrev4-1} % PRD
%\bibliographystyle{model1-num-names}  % PLB with title + plb.bst파일 같은 디렉토리에 첨부
\bibliographystyle{JHEP}       %% JHEP.bst+ jhep.bst파일 같은 디렉토리에 첨부

\bibliography{references}

\end{document}